\documentclass{article}

\PassOptionsToPackage{numbers, compress}{natbib}

\usepackage[preprint]{neurips_2023}

\usepackage{amsmath,amsfonts,bm}

\def\eqref#1{equation~\ref{#1}}

\def\1{\bm{1}}

\def\rr{{\textnormal{r}}}

\def\rx{{\textnormal{x}}}
\def\ry{{\textnormal{y}}}
\def\rz{{\textnormal{z}}}

\DeclareMathAlphabet{\mathsfit}{\encodingdefault}{\sfdefault}{m}{sl}
\SetMathAlphabet{\mathsfit}{bold}{\encodingdefault}{\sfdefault}{bx}{n}

\def\sB{{\mathbb{B}}}

\usepackage[utf8]{inputenc} 
\usepackage[T1]{fontenc}    
\usepackage{hyperref}       
\usepackage{url}            
\usepackage{booktabs}       
\usepackage{amsfonts}       
\usepackage{nicefrac}       
\usepackage{microtype}      
\usepackage{xcolor}         

\usepackage{amsmath}
\usepackage{amssymb}
\usepackage{multirow}

\usepackage{graphicx}
\usepackage{wrapfig}

\usepackage{makecell}

\usepackage{algorithm}  
\usepackage{algorithmicx}  
\usepackage{algpseudocode} 
\usepackage{paralist}
\usepackage{bbm}

\usepackage[capitalize]{cleveref}
\crefname{section}{Sec.}{Secs.}
\Crefname{section}{Section}{Sections}
\Crefname{table}{Table}{Tables}
\crefname{table}{Tab.}{Tabs.}

\newtheorem{theorem}{Theorem}
\newtheorem{corollary}{Corollary}
\newtheorem{proof}{Proof}

\title{PrivaScissors: Enhance the Privacy of Collaborative Inference through the Lens of Mutual Information}

\author{%
  Lin Duan\thanks{Both authors contributed equally to this research and are placed according to alphabetical order.}, Jingwei Sun\footnotemark[1], Yiran Chen, Maria Gorlatova \\
  Department of Electrical and Computer Engineering\\
  Duke University\\
  \texttt{\{lin.duan, jingwei.sun, yiran.chen, maria.gorlatova\}@duke.edu} \\
}

\begin{document}

\maketitle

\begin{abstract}
  Edge-cloud collaborative inference empowers resource-limited IoT devices to support deep learning applications without disclosing their raw data to the cloud server, thus preserving privacy. Nevertheless, prior research has shown that collaborative inference still results in the exposure of data and predictions from edge devices. To enhance the privacy of collaborative inference, we introduce a defense strategy called \textit{\textbf{PrivaScissors}}, which is designed to reduce the mutual information between a model's intermediate outcomes and the device's data and predictions. We evaluate PrivaScissors's performance on several datasets in the context of diverse attacks and offer a theoretical robustness guarantee.
\end{abstract}

\section{Introduction}


Edge devices are rapidly evolving, becoming smarter and more versatile. These devices are expected to perform a wide range of deep learning (DL) inference tasks with high efficiency and remarkable performance. However, implementing DL inference applications on such edge devices can be quite challenging due to the constraints imposed by the on-device resource availability. As we see the rise of state-of-the-art DL models, such as Large Language Models, they are becoming increasingly complex, housing a colossal number of parameters. This escalation in complexity and size makes it difficult to store a DL model on an edge device, which typically has limited memory space. Furthermore, the restricted computational resources could lead to unacceptably long latency during inference. One potential solution to this predicament is to transmit the input data directly from the edge device to a cloud server. The server, which houses the DL model, then conducts inference and sends the prediction back to the device. However, this approach carries with it the risk of privacy breaches, particularly if the input data are sensitive in nature - such as facial images. It's also important to note that the predictions can also contain confidential information, such as the patient's diagnostic results.

\begin{figure}[th]
\centering
     \includegraphics[scale=0.35]{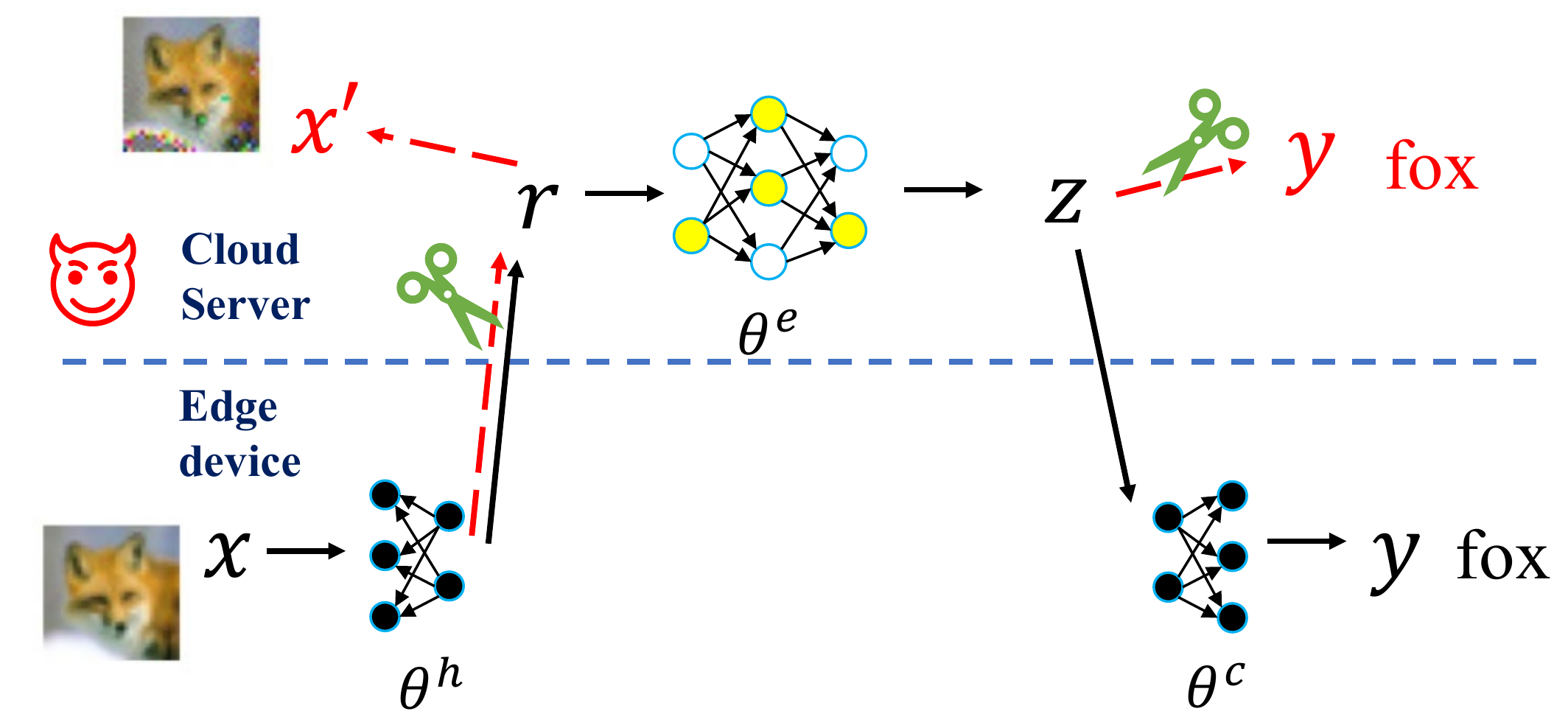}
\caption{A general framework of collaborative inference. The malicious server can infer data and predictions on the edge device. PrivaScissors defends against privacy leakage by reducing the mutual information between the model's intermediate outcomes and the edge device's data and predictions.}
\label{fig:overview}
\end{figure}

Collaborative inference~\cite{li2018auto,shlezinger2021collaborative,zhou2021bbnet} has become a privacy-preserving approach to deploying DL inference applications on commodity edge devices that have limited computing resources. \cref{fig:overview} shows a general collaborative inference system. Suppose an edge device and a cloud server conduct collaborative inference. The deep learning model can be divided into three parts\footnote{It is notable that some applications might divide the model into two parts, and the edge devices might hold the first or the last few layers, which have different privacy leakage problems. This paper considers the general setting, which has privacy concerns in both settings.}. The first and last few layers of the network are deployed on the edge device, while the remaining layers are offloaded to a cloud server. This division allows most of the computational tasks to be handled by the server, effectively mitigating the resource limitations on the device. The edge device and the cloud server communicate only the intermediate outputs of the model, ensuring that the raw data and predictions remain inaccessible to the server. However, recent works~\cite{he2019model,he2020attacking} have revealed that sharing these intermediate outputs can still lead to data and prediction leakage. A malicious server can, for instance, reconstruct input data from the representations (i.e., $r$ in \cref{fig:overview}) uploaded by the device through Model Inversion (MI) attacks~\cite{zhu2019deep, zhao2020idlg,he2020attacking}. Furthermore, the high-level features (i.e., $z$ in \cref{fig:overview}) contain rich information about the predictions, making it feasible for a malicious server to infer the device's predictions through these features~\cite{fu2022label, li2021label, liu2021defending}. While there have been considerable explorations into privacy preservation in collaborative inference~\cite{he2019model,he2020attacking,wang2021improving,zou2023mutual}, existing defenses tend to significantly degrade model utility. This degradation is particularly evident in scenarios where attacks are relatively strong.
For example, when the head model on the device (i.e., $\theta^h$ in \cref{fig:overview}) is shallow, existing defenses cannot guarantee the privacy of the input data without a significant drop in model accuracy.

We propose a defense method named \textit{PrivaScissors}, designed from a mutual information perspective to enhance the edge device's privacy in collaborative inference. This approach works by protecting both the device's data and its predictions. To protect the device's data, we regularize the head model on the device to extract representations that hold less mutual information with the input. To protect the prediction, we regularize the features extracted by the server's encoder to minimize the mutual information they contain with the label. We derive a variational mutual information upper-bound and develop an adversarial training method to minimize this bound on the device side. Our defense's robustness is theoretically guaranteed. We evaluate PrivaScissors on CIFAR10 and CIFAR100 against both black-box and white-box MI attacks. The results show that PrivaScissors can effectively defend the attacks with less than a 3\% drop in model accuracy even when the head model on the device has only one convolutional layer, where the attacks are extremely strong. We also evaluate our defense against prediction leakage using multiple model completion (MC) attacks~\cite{fu2022label,li2021label}. The results show that our defense achieves the best trade-off between the model accuracy and defending effectiveness compared with the baselines.

Our contributions are summarized as follows:

\begin{itemize}
    \item To the best of our knowledge, this is the first paper to systematically address the privacy leakage in collaborative inference, encompassing both data leakage and prediction leakage.
    \item We propose a defense method against data and prediction leakage in collaborative inference from the mutual information perspective. We offer a theoretical robustness guarantee of our defense against general privacy leakage from the intermediate outcomes of a model.
    \item We empirically evaluate our defense across multiple datasets and against multiple attacks. The results show that our defense can defend MI attacks while preserving high accuracy, even when the head model has only one convolutional layer. Our defense can also prevent prediction leakage against MC attacks with nearly no model accuracy drop.
\end{itemize}
\section{Related Work}

\subsection{Privacy Leakage in Collaborative Inference}

Privacy leakage is drawing more and more attention as the rapid growth of commercial deployment of deep learning, especially in collaborative learning scenarios, whose primary concern is privacy. 
In collaborative inference, we categorize privacy leakage into two types, i.e., data leakage~\cite{luo2021feature, he2019model, jiang2022comprehensive, jin2021cafe} and prediction leakage~\cite{fu2022label, li2021label, liu2021defending}. 

For data leakage, \cite{luo2021feature} proposes general attack methods for complex models, such as Neural Networks, by matching the correlation between adversary features and target features, which can be seen as a variant of model inversion~\cite{fredrikson2015model,sun2021soteria}. \cite{he2019model, geiping2020inverting, jin2021cafe, yin2021see, melis2019exploiting, jiang2022comprehensive} also propose variants of model inversion attack. While all these attacks are in the inference phase, \cite{jin2021cafe} proposes a variant of DLG~\cite{zhu2019deep} which can perform attacks in the training phase. For prediction leakage, \cite{li2021label} proposes an attack and defense method for two-party split learning on binary classification problems, a special collaborative inference setting. Additionally, \cite{fu2022label} proposes three different label inference attack methods considering different settings in collaborative inference: direct label inference attack, passive label inference attack, and active label inference attack. 

\subsection{Defense in Collaborative Inference}

Defensive methods have also been proposed against privacy leakage in collaborative inference. To defend against data leakage, some works apply differential privacy (DP)~\cite{he2019model,he2020attacking} and compression~\cite{he2019model,he2020attacking,wang2021improving} to the representations and models. These methods can sometimes defend against data leakage from the representation, but they also cause substantial model performance degradation because they destroy the knowledge/information in the representations. Two recent works also try to solve the data leakage problem from the mutual information perspective~\cite{wang2021improving,zou2023mutual}. However, their methods only achieve decent results when the head model on the edge device is deep, which is not practical when the computation power is constrained on the edge device. To defend against prediction leakage, \cite{liu2021defending} manipulates the labels following specific rules to defend the direct label inference attack, which can be seen as a variant of label differential privacy (label DP)~\cite{chaudhuri2011sample, ghazi2021deep} in collaborative inference. Compression and quantization of the gradients~\cite{fu2022label, zou2023mutual} are also applied to defend against prediction leakage. However, similarly to the defense against data leakage, these defenses cause substantial model performance degradation to achieve decent defending performance.
\section{Preliminary}

\subsection{Collaborative Inference Setting}\label{sec:setup}

Suppose an edge device and a cloud server conduct collaborative inference. Following the setting in \cref{fig:overview}, the deep learning model is divided into a head model $f^h_{\theta^h}$, an encoder $f^e_{\theta^e}$ and a classifier $f^c_{\theta^c}$. The head model and classifier are deployed on the edge device, and the encoder is on the cloud server. Given an input $x_i$, the edge device first calculates the representation $r_i=f^h_{\theta^h}(x_i)$ and sends $r_i$ to the server. Then the server extracts the feature from the received representation $z_i=f^e_{\theta^e}(r_i)$ and sends the feature back to the edge device. After receiving the feature, the edge device calculates the prediction $\hat{y}_i=f^c_{\theta^c}(z_i)$. In this paper, the results of $f^h_{\theta^h}$ sent from the device to the server are referred to as \textit{representations}, and \textit{features} refer to the results of $f^e_{\theta^e}$ sent from the server to the device. The overall inference procedure can be formulated as

\begin{equation}
    \hat{y}_i = f^c_{\theta^c}(f^e_{\theta^e}(f^h_{\theta^h}(x_i))).
\end{equation}




In the real world, the raw data $x_i$ and prediction $\hat{y}_i$ are important intelligent properties of the edge device and may contain private information. In the inference procedure, the edge device does not send raw data to the server, and the inference results are also inaccessible to the server. 

\subsection{Threat Model}

We consider that the edge device possessing the head model and the classifier is trusted. The edge device only uploads the representations to the server and never leaks raw data or predictions to the server. However, the cloud server is untrusted, attempting to steal raw data and predictions. We assume the untrusted server strictly follows the collaborative inference protocols, and it cannot compromise the inference process conducted by the edge device. With the received representation $r_i$, the server can reconstruct the input data $x_i$ on the edge device by conducting model inversion (MI) attacks~\cite{zhu2019deep, zhao2020idlg,he2019model}. Notably, the head model on the edge device is usually shallow due to the computation resource limitation, which aggravates data leakage vulnerability from the representation~\cite{he2020attacking}. The encoder on the server extracts high-level features containing rich information about the prediction, which enables the server to infer predictions of the device.

\section{Method}
\subsection{Defense Formulation}\label{sec:defense_form}

To defend against privacy leakage, we propose a learning algorithm that regularizes the model during the training phase. Following the setup of \ref{sec:setup}, suppose the edge device has sample pairs $\left\{\left(x_i, y_i\right)\right\}_{i=1}^N$ drawn from a distribution $p\left(\rx, \ry\right)$. The representation is calculated as $\rr=f^h_{\theta^h}(\rx)$ by the edge device, and the cloud server computes features $\rz=f^e_{\theta^e}(\rr)$. We apply $\rx, \ry, \rr, \rz$ here to represent random variables, while $x_i, y_i, r_i, z_i$ are deterministic values. To preserve the privacy of the edge device's raw data and inference results, our learning algorithm is to achieve three goals:

\begin{itemize}
    \item Goal 1: To preserve the performance of collaborative inference, the main objective loss should be minimized.
    \item Goal 2: To reduce the data leakage from the representations, $\theta^h$ should not extract representations $\rr$ containing much information about the raw data $\rx$.
    \item Goal 3: To reduce the leakage of the predictions on the edge device, $\theta^e$ on the cloud server should not be able to extract features $\rz$ containing much information about the true label $\ry$.
\end{itemize}

\noindent Formally, we have three training objectives:

\begin{equation}
    \begin{aligned}
    &\textbf{Prediction:} \min\limits_{\theta^h, \theta^e, \theta^c} \mathcal{L}\left(f^c_{\theta^c}\left(f^e_{\theta^e}\left(f^h_{\theta^h}(\rx)\right)\right), \ry\right),\\
    &\textbf{Data protection:} \min\limits_{\theta^h} \text{I}(\rr;\rx),\\
    &\textbf{Prediction protection:} \min\limits_{\theta^h, \theta^e} \text{I}(\rz;\ry),
    \end{aligned}
\end{equation}

where $\text{I}(\rr;\rx)$ is the mutual information between the representation and the data, which indicates how much information $\rr$ preserves for the data $\rx$. Similarly, $\text{I}(\rz;\ry)$ is the mutual information between the feature and the label. We minimize these mutual information terms to prevent the cloud server from inferring the data $\rx$ and label $\ry$ from $\rr$ and $\rz$, respectively.

The prediction objective is usually easy to optimize (e.g., cross-entropy loss for classification). However, the mutual information terms are hard to calculate in practice for two reasons: 1. $\rr$ and $\rx$ are high-dimensional, and it is extremely computationally heavy to compute their joint distribution; 2. Calculating the mutual information requires knowing the distributions p(x|r) and p(y|z), which are both difficult to compute. To derive tractable estimations of the mutual information objectives, we leverage CLUB\cite{cheng2020club} to formulate variational upper-bounds of mutual information terms. We first formulate a variational upper-bound of $\text{I}(\rr;\rx)$:

{\small
\begin{equation}
    \begin{aligned}
        &\text{I}\left(\rr;\rx\right)\\
        \leq~& \text{I}_\text{{vCLUB}}\left(\rr;\rx\right)\\
        :=~& \mathbb{E}_{p\left(\rr, \rx\right)}\log q_{\psi}\left(\rx|\rr\right)-\mathbb{E}_{p\left(\rr\right)p\left(\rx\right)}\log q_{\psi}\left(\rx|\rr\right),
    \end{aligned}
    \label{eq:vclub}
\end{equation}
}%

where $q_{\psi}\left(\rx|\rr\right)$ is a variational distribution with parameters $\psi$ to approximate $p\left(\rx|\rr\right)$. To guarantee the inequality of \cref{eq:vclub}, $q_{\psi}\left(\rx|\rr\right)$ should satisfy
{\small
\begin{equation}
    \text{KL}\left(p\left(\rr,\rx\right)||q_\psi\left(\rr,\rx\right)\right)\leq \text{KL}\left(p\left(\rr\right)p\left(\rx\right)||q_\psi\left(\rr,\rx\right)\right),
\end{equation}
}%

which can be achieved by minimizing $\text{KL}\left(p\left(\rr,\rx\right)||q_\psi\left(\rr,\rx\right)\right)$:
{\small
\begin{equation}
    \begin{aligned}
        \psi~=~&\arg\min\limits_\psi \text{KL}\left(p\left(\rr,\rx\right)||q_\psi\left(\rr,\rx\right)\right)\\
        =~&\arg\min\limits_\psi \mathbb{E}_ {p\left(\rr,\rx\right)}\left[\log\left(p\left(\rx|\rr\right)p\left(\rr\right)\right)-\log\left(q_{\psi}\left(\rx|\rr\right)p\left(\rr\right)\right)\right]\\
        =~&\arg\max\limits_\psi \mathbb{E}_ {p\left(\rr,\rx\right)}\log\left(q_{\psi}\left(\rx|\rr\right)\right).
    \end{aligned}\label{eq:condition}
\end{equation}
}%

With sample pairs $\left\{\left(x_i, y_i\right)\right\}_{i=1}^N$, we can apply the sampled vCLUB (vCLUB-S) mutual information estimator in \cite{cheng2020club} to reduce the computational overhead, which is an unbiased estimator of $\text{I}_\text{{vCLUB}}$ and is formulated as
{\small
\begin{equation}
    \hat{\text{I}}_{\text{vCLUB-S}}(\rr;\rx) = \frac{1}{N} \sum\limits_{i=1}^N \left[\log q_{\psi}\left(x_i|r_i\right) - \log q_{\psi}\left(x_{k_i'}|r_i\right)\right],
\label{eq:vclubs}
\end{equation}
}%

where $k_i'$ is uniformly sampled from indices $\{1,...,N\}$. With \cref{eq:vclub}, \cref{eq:condition} and \cref{eq:vclubs}, the objective of data protection is formulated as:
{\small
\begin{equation}
    \begin{aligned}
        &\min\limits_{\theta^h} \text{I}(\rr;\rx)  \Leftrightarrow~\min\limits_{\theta^h} \hat{\text{I}}_{\text{vCLUB-S}}(\rr;\rx)\\
        =~&\min\limits_{\theta^h} \frac{1}{N} \sum\limits_{i=1}^N \left[\max\limits_{\psi}\log q_{\psi}\left(x_i|r_i\right) - \log q_{\psi}\left(x_{k_i'}|r_i\right)\right].
    \end{aligned}
    \label{eq:variational_objective_data}
\end{equation}
}%

Similarly, we can use a variational distribution $q_{\phi}(y|z)$ with parameter $\phi$ to approximate $p(y|z)$, and formulate the objective of label protection as:
{\small
\begin{equation}
    \begin{aligned}
        &\min\limits_{\theta^h, \theta^e} \text{I}(\rz;\ry)\Leftrightarrow~\min\limits_{\theta^h, \theta^e} \hat{\text{I}}_{\text{vCLUB-S}}(\rz;\ry)\\        =~&\min\limits_{\theta^h, \theta^e} \frac{1}{N} \sum\limits_{i=1}^N \left[\max\limits_{\phi}\log q_{\phi}\left(y_i|z_i\right) - \log q_{\phi}\left(y_{n_i'}|z_i\right)\right].
    \end{aligned}
    \label{eq:variational_objective_label}
\end{equation}
}%

Suppose we use $g_{\psi}$, $h_{\phi}$ to parameterize $q_{\psi}$ and $q_{\phi}$, respectively. By combining \cref{eq:variational_objective_data}, \cref{eq:variational_objective_label} and the prediction objective with weight hyper-parameters $\lambda_d$ and $\lambda_l$, we formulate the overall optimizing objective as

{\small
\begin{equation}
    \begin{aligned}
        &\min\limits_{\theta^h, \theta^e, \theta^c} (1-\lambda_d-\lambda_l)\underbrace{\frac{1}{N} \sum\limits_{i=1}^N\mathcal{L}\left(f^c_{\theta^c}\left(f^e_{\theta^e}\left(f^h_{\theta^h}(x_i)\right)\right), y_i\right)}_{\mathcal{L}_c}\\
        &+\min\limits_{\theta^h}\max\limits_{\psi} \lambda_d\underbrace{\frac{1}{N} \sum\limits_{i=1}^N \log g_{\psi}\left(x_i|f^h_{\theta^h}(x_i)\right)}_{\mathcal{L}_{d\_a}} +\min\limits_{\theta^h} \lambda_d\underbrace{\frac{1}{N} \sum\limits_{i=1}^N-\log g_{\psi}\left(x_{k_i'}|f^h_{\theta^h}(x_i)\right)}_{\mathcal{L}_{d\_r}}\\
        &+\min\limits_{\theta^h, \theta^e}\max\limits_{\phi} \lambda_l\underbrace{\frac{1}{N} \sum\limits_{i=1}^N \log h_{\phi}\left(y_i|f^e_{\theta^e}\left(f^h_{\theta^h}(x_i)\right)\right)}_{\mathcal{L}_{l\_a}} +\min\limits_{\theta^h, \theta^e} \lambda_l\underbrace{\frac{1}{N} \sum\limits_{i=1}^N-\log h_{\phi}\left(y_{n_i'}|f^e_{\theta^e}\left(f^h_{\theta^h}(x_i)\right)\right)}_{\mathcal{L}_{l\_r}}.
        \label{eq:final_object}
    \end{aligned}
\end{equation}
}%

$h_{\phi}$ can be easily constructed to estimate $p\left(\ry|\rz\right)$ given the task of inference (e.g., classifier for classification task). To estimate $p\left(\rx|\rr\right)$, we assume that $\rx$ follows the Gaussian distribution of which the mean vector is determined by $\rr$ and the variance is 1. Under this assumption, we apply a generator $g_{\psi}$ to estimate the mean vector of $\rx$ given $\rr$.

\begin{figure}[th]
\centering
     \includegraphics[scale=0.45]{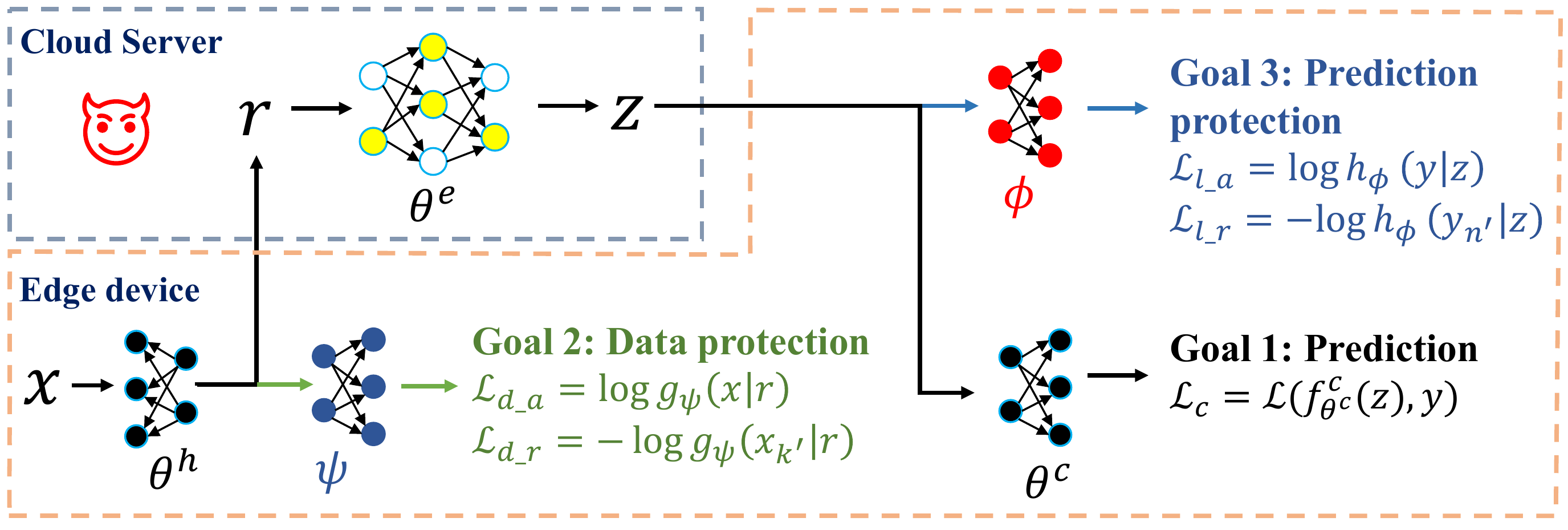}
\caption{An overview of PrivaScissors. Training step 1: Optimize the classifiers $\theta^c$ and $\phi$ by minimizing $\mathcal{L}_c$ and maximizing $\mathcal{L}_{l\_a}$, respectively. Step 2: Optimize the generator $\psi$ by maximizing $\mathcal{L}_{d\_a}$. Step 3: Optimize $\theta^h$ and $\theta_e$ by minimizing $(1-\lambda_d-\lambda_l)\mathcal{L}_c+\lambda_l\mathcal{L}_{l\_a}+\lambda_l\mathcal{L}_{l\_r}+\lambda_d\mathcal{L}_{d\_a}+\lambda_d\mathcal{L}_{d\_r}$.}
\label{fig:framework}
\end{figure}

\subsection{Learning Algorithm}

The overall objective has five terms. For simplicity, we denote these five objective terms as $\mathcal{L}_c$, $\mathcal{L}_{d\_a}$, $\mathcal{L}_{d\_r}$, $\mathcal{L}_{l\_a}$ and $\mathcal{L}_{l\_r}$, respectively, as shown in \cref{eq:final_object}. $\mathcal{L}_c$ is the prediction objective. $\mathcal{L}_{d\_a}$ and $\mathcal{L}_{d\_r}$ comprise the data protection objective. $\mathcal{L}_{d\_a}$ is an adversarial training objective where an auxiliary generator $g_{\psi}$ is trained to capture data information while the head layers $f^{h}_{\theta^h}$ are trained to extract as little data information as possible. $\mathcal{L}_{d\_r}$ regularizes $f^{h}_{\theta^h}$ to extract representations that can be used to generate randomly picked samples. $\mathcal{L}_{l\_a}$ and $\mathcal{L}_{l\_r}$ have similar effect with $\mathcal{L}_{d\_a}$ and $\mathcal{L}_{d\_r}$, respectively. We can reorganize the overall training objective as

{\small
\begin{equation}
\begin{aligned}
    &\theta^h, \theta^e, \theta^c, \psi, \phi
    =~\arg\min\limits_{\theta^h,\theta^e}\left[(1-\lambda_d-\lambda_l)\min\limits_{\theta^c}\mathcal{L}_c+\lambda_l\max\limits_{\phi}\mathcal{L}_{l\_a}+\lambda_l\mathcal{L}_{l\_r}+\lambda_d\max\limits_{\psi}\mathcal{L}_{d\_a}+\lambda_d\mathcal{L}_{d\_r}\right].
\end{aligned}\label{eq:organized_object}
\end{equation}
}%

Based on \cref{eq:organized_object}, we develop a collaborative learning algorithm. For each batch of data, the device first optimizes the classifiers $\theta^c$ and $\phi$ by minimizing $\mathcal{L}_c$ and maximizing $\mathcal{L}_{l\_a}$, respectively. Then, the device optimizes the generator $\psi$ by maximizing $\mathcal{L}_{d\_a}$. Finally, $\theta^h$ and $\theta_e$ are optimized by minimizing $(1-\lambda_d-\lambda_l)\mathcal{L}_c+\lambda_l\mathcal{L}_{l\_a}+\lambda_l\mathcal{L}_{l\_r}+\lambda_d\mathcal{L}_{d\_a}+\lambda_d\mathcal{L}_{d\_r}$. The detailed algorithm can be found in \cref{app:algorithm}. Note that $\theta^h, \theta^c, \psi$ and $\phi$ are deployed on devices, and their training does not need additional information from the cloud server compared with training without our defense. The training procedure of $\theta^e$ does not change, which makes our defense concealed from the cloud server.

\subsection{Robustness Guarantee}\label{sec:guarantee}

We derive certified robustness guarantees for our defenses against prediction and data leakage. Following the notations in \cref{sec:defense_form}, we have the following theorem of robustness guarantee for prediction leakage after applying PrivaScissors. All the proofs can be found in \cref{app:proof}.

\begin{theorem}
\label{thm:prediction}
Let $h_{\phi}$ parameterize $q_{\phi}$ in \cref{eq:variational_objective_label}. Suppose the malicious server optimizes an auxiliary model $h^m(\ry|\rz)$ to estimate $p(\ry|\rz)$. For any $h^m(\ry|\rz)$, we always have:

{\small
\begin{equation}
    \frac{1}{N}\sum\limits_{i=1}^N \log h^m(y_i|z_i) < \frac{1}{N}\sum\limits_{i=1}^N \log p(y_i) + \epsilon,
\end{equation}
}%

where 

{\small
\begin{equation}
    \epsilon = \text{I}_{\text{vCLUB}_{h_\phi}}(\rz;\ry) + \text{KL}(p(\ry|\rz)||h_{\phi}(\ry|\rz)).
\end{equation}
}%

\end{theorem}

Specifically, if the task of collaborative inference is classification, we have the following corollary:

\begin{corollary}
Suppose the task of collaborative inference is classification. Following the notations in \cref{thm:prediction} and let $epsilon$ be defined therein, we have:

{\small
\begin{equation}
    \frac{1}{N}\sum\limits_{i=1}^N\text{CE}\left[h^m(z_i), y_i\right] > CE_{random} - \epsilon,
\end{equation}
}%

where CE denotes the cross-entropy loss, and $\text{CE}_{random}$ is the cross-entropy loss of random guessing.

\end{corollary}

For data leakage, we have the following theorem of robustness.

\begin{theorem}
\label{thm:data}
Let the assumption of $p(\rx|\rr)$ in \cref{sec:defense_form} hold and $g_{\psi}$ parameterize the mean of $q_{\psi}$ in \cref{eq:variational_objective_data}. $Q$ denotes the dimension of $\rx$. Suppose the malicious server optimizes an auxiliary model $g^m(\rx|\rr)$ to estimate the mean of $p(\rx|\rr)$. For any $g^m(\rx|\rr)$, we always have:

{\small
\begin{equation}
    \frac{1}{N}\sum\limits_{i=1}^N \text{MSE}\left[g^m(r_i), x_i\right] > \frac{2(\kappa-\epsilon)}{Q},
\end{equation}
}%

where MSE denotes the \textbf{mean square error}, and

{\small
\begin{equation}
\begin{aligned}
    &\kappa = -\frac{1}{N}\sum\limits_{i=1}^N \log \frac{\sqrt{2\pi}}{p(x_i)},\\
    &\epsilon = \text{I}_{\text{vCLUB}_{g_\psi}}(\rr;\rx) + \text{KL}(p(\rx|\rr)||g_{\psi}(\rx|\rr)).
\end{aligned}
\end{equation}
}%

\end{theorem}

\section{Experiments}\label{sec:experiments}
We first evaluate PrivaScissors against data leakage and prediction leakage separately. Then we evaluate the integration of defenses against data and prediction leakages.

\subsection{Experimental Setup}

\paragraph{Attack methods}
For data leakage, we evaluate PrivaScissors against two model inversion (MI) attacks: (1) \textbf{Knowledge Alignment (KA)}~\cite{wang2021improving}  is a black-box MI attack, in which the malicious server trains an inversion model that swaps the input and output of the target model using an auxiliary dataset. The inversion model is then used to reconstruct the input given any representation. (2) \textbf{Regularized Maximum Likelihood Estimation (rMLE)}~\cite{he2020attacking} is a white-box MI attack that the malicious server has access to the device's extractor model $\theta^h$. The server trains input to minimize the distance between the fake representations and the received ground-truth representations. It is an unrealistic assumption that the server can access the model on the device, and we apply this white-box attack to evaluate our defense against extremely strong attacks. For prediction leakage, we evaluate our defense against two attacks: (1) \textbf{Passive Model Completion (PMC)}~\cite{fu2022label} attack assumes that the malicious server has access to an auxiliary labeled dataset and utilizes this auxiliary dataset to fine-tune a classifier that can be applied to its encoder. (2) \textbf{Active Model Completion (AMC)}~\cite{fu2022label} attack is conducted by the server to trick the collaborative model into relying more on its feature encoder.

\vspace{-2mm}
\paragraph{Baselines}

We compare PrivaScissors with four existing defense baselines: (1) \textbf{Adding Noise (AN)}~\cite{chong2022baseline} is proven effective against privacy leakage in collaborative learning by adding Laplacian noise to the representations and gradients. (2) \textbf{Data Compression (DC)}~\cite{chong2022baseline} prunes representations and gradients that are below a threshold magnitude, such that only a part of the representations and gradients are sent to the server. (3) \textbf{Privacy-preserving Deep Learning (PPDL)}~\cite{shokri2015ppdl} is a comprehensive privacy-enhancing method including three defense strategies: differential privacy, data compression, and random selection. (4) \textbf{Mutual Information Regularization Defense (MID)}~\cite{zou2023mutual} is the SOTA defense against privacy leakage in split learning and collaborative inference. MID is also based on mutual information regularization by applying \textit{Variational Information Bottleneck (VIB)}.

\vspace{-2mm}
\paragraph{Dataset \& Hyperparameter configurations} We evaluate on CIFAR10 and CIFAR100. For both datasets, we apply ResNet18 as the backbone model. The first convolutional layer and the last basic block are deployed on the device as the representation extractor and the classifier, respectively. We set batch size $B$ as 32 for both datasets. We apply SGD as the optimizer with the learning rate $\eta$ set to be 0.01. 
The server has 40 and 400 labeled samples to conduct KA and MC attacks for CIFAR10 and CIFAR100, respectively.
For PrivaScissors, we apply a 1-layer decoder and a 3-layer MLP to parameterize $\psi$ and $\phi$. For AN defense, we apply Laplacian noise with mean of zero and scale between 0.0001-0.01. For DC baseline, we set the compression rate from 90\% to 100\%. For PPDL, we set the Laplacian noise with scale of 0.0001-0.01, $\tau=0.001$ and $\theta$ between 0 and 0.01. For MID baseline, we set the weight of mutual information regularization between 0-0.1.

\vspace{-2mm}
\paragraph{Evaluation metrics} (1) \textbf{Utility metric (Model accuracy)}: We use the test data accuracy of the classifier on the device to measure the performance of the collaborative model. (2) \textbf{Robustness metric (SSIM)}: We use SSIM (structural similarity) between the reconstructed images and the raw images to evaluate the effectiveness of defense against data leakage. The lower the SSIM, the better the defense performance. (3) \textbf{Robustness metric (Attack accuracy)}: We use the test accuracy of the server’s classifier after conducting MC attacks to evaluate the defense against prediction leakage. The lower the attack accuracy, the higher the robustness against prediction leakage.

\subsection{Results of Data Protection} \label{sec:results_data}

\begin{figure}[th]
\centering
     \includegraphics[scale=0.34]{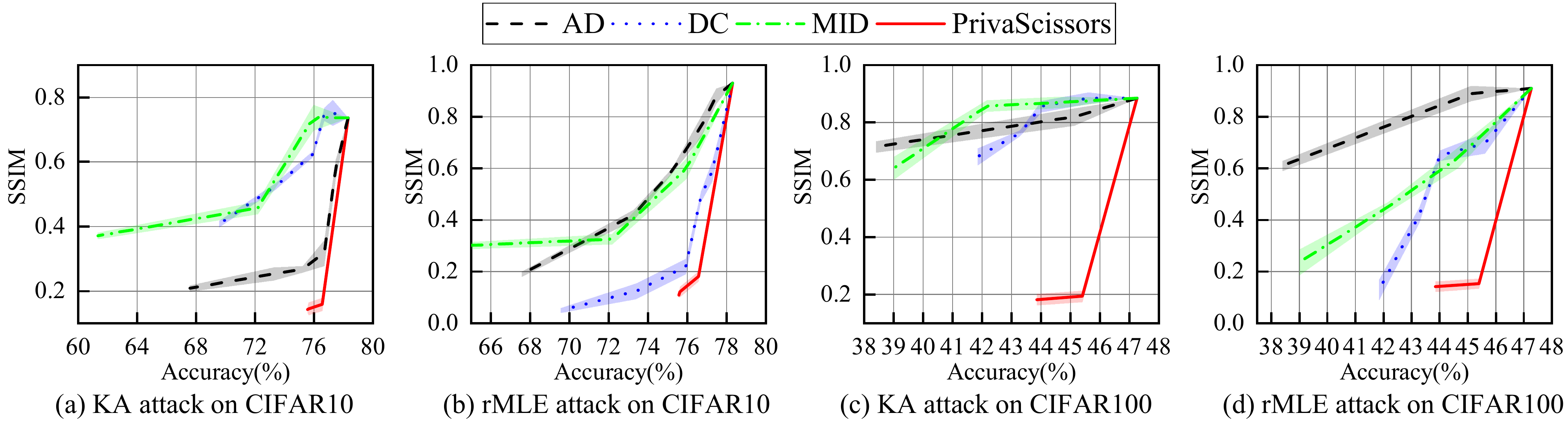}
\caption{Results of model accuracy v.s. SSIM of reconstruction on CIFAR10 and CIFAR100 against rMLE and KA attack.}
\label{fig:MI_tradeoff}
\end{figure}

We conduct experiments on CIFAR10 and CIFAR100 to evaluate our defense against the KA attack and the rMLE attack. We set different defense levels for our methods (i.e., different $\lambda_d$ values in \cref{eq:final_object}) and baselines to conduct multiple experiments to show the trade-off between the model accuracy and SSIM of reconstruction. The results are shown in \cref{fig:MI_tradeoff}.

For defense against KA attacks, our PrivaScissors can reduce the SSIM of reconstruction to lower than 0.2 with a model accuracy drop of less than 2\% for CIFAR10. In contrast, the other baselines drop model accuracy by more than 10\% and cannot achieve the same defense effect even with an accuracy drop of more than 10\%. Notably, the malicious server has more auxiliary data on CIFAR100 than CIFAR10, making the attack harder to defend on CIFAR100 to the baselines. However, PrivaScissors can still achieve an SSIM of lower than 0.2 with a model accuracy drop of less than 2\%. We also evaluate our defense against the KA attack with a larger auxiliary dataset on the malicious server, and the results show that our defense can effectively defend against the KA attack when the server has more auxiliary samples. For defense against rMLE attacks, PrivaScissors achieves similar results of reducing the SSIM to lower than 0.2 with a model accuracy drop of less than 2\% for CIFAR10 and 1\% for CIFAR100, respectively, which outperforms the other baselines significantly. 

\begin{figure}[th]
\centering
     \includegraphics[scale=0.4]{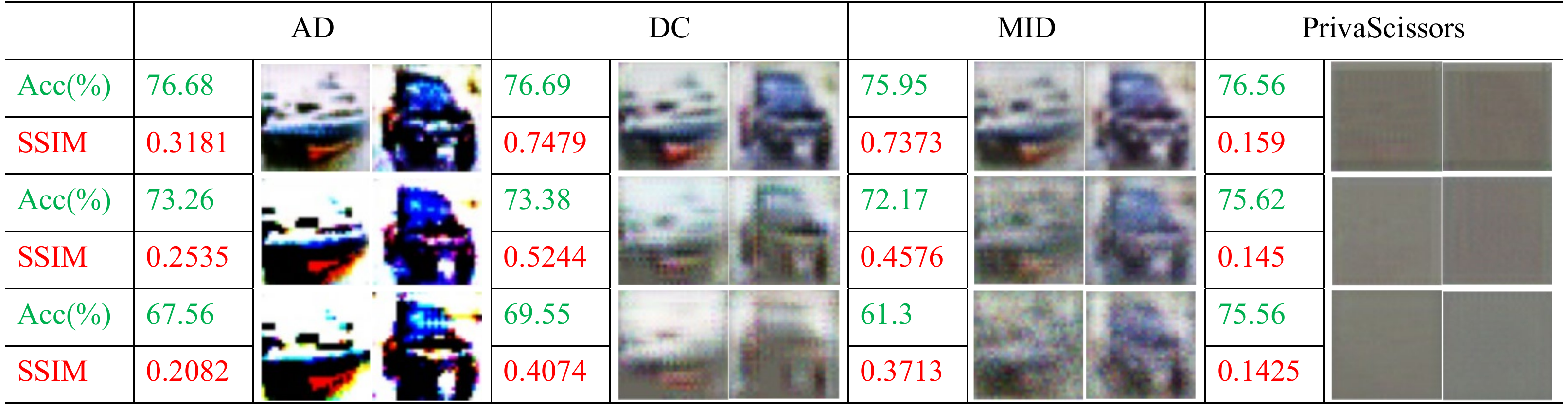}
\caption{Images reconstructed by the KA attack on CIFAR10 under different defenses.}
\label{fig:MI_attack}
\end{figure}

To perceptually demonstrate the effectiveness of our defense, we show the reconstructed images by the KA attack on CIFAR10 after applying baseline defenses and our defense in \cref{fig:MI_attack}. It is shown that by applying the baseline defenses, the reconstructed images still contain enough information to be recognizable with the model accuracy of lower than 70\%. For our method, the reconstructed images do not contain much information about the raw images, with the model accuracy higher than 76\%.

\subsection{Results of Prediction Protection} \label{sec:results_prediction}

\begin{figure}[th]
\centering
     \includegraphics[scale=0.34]{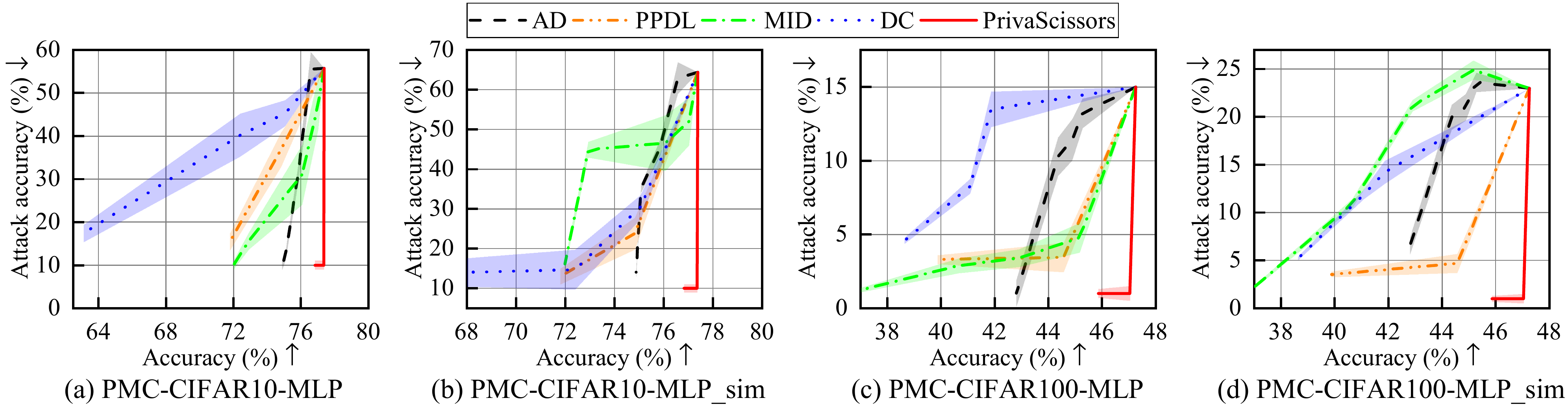}
\caption{Results of model accuracy v.s. attack accuracy on CIFAR10 and CIFAR100 against PMC attack.}
\label{fig:PMC}
\end{figure}

\begin{figure}[th]
\centering
     \includegraphics[scale=0.34]{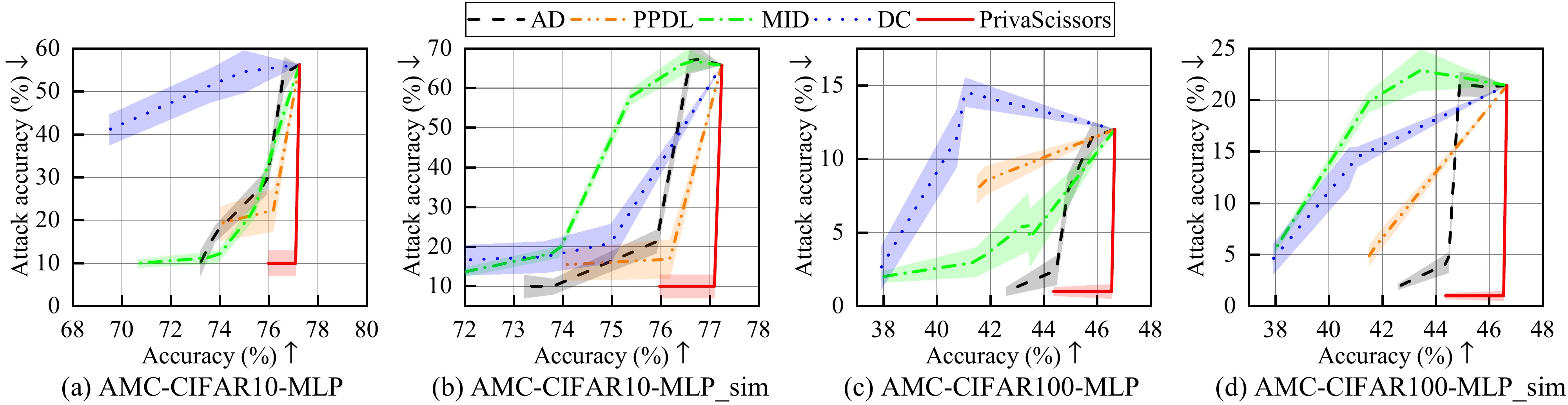}
\caption{Results of model accuracy v.s. attack accuracy on CIFAR10 and CIFAR100 against AMC attack.}
\label{fig:AMC}
\end{figure}

We evaluate PrivaScissors on two datasets against two attack methods. We set different defense levels for our methods (i.e., different $\lambda_l$ values in \cref{eq:final_object}) and baselines to conduct multiple experiments to show the trade-off between the model accuracy and attack accuracy. The defense results against PMC and AMC attacks are shown in \cref{fig:PMC} and \cref{fig:AMC}, respectively. To simulate the realistic settings in that the malicious server uses different model architectures to conduct MC attacks, we apply different model architectures (MLP \& MLP\_sim) for MC attacks. 

For defense against PMC on CIFAR10, PrivaScissors achieves 10\% attack accuracy (equal to random guess) by sacrificing less than 0.5\% model accuracy, while the other baselines suffer a model accuracy drop by more than 4\% to achieve the same defense effect. Similarly, PrivaScissors achieves 1\% attack accuracy on CIFAR100 by sacrificing less than 1\% model accuracy, while the other baselines achieve the same defense effect by sacrificing more than 6\% model accuracy.

PrivaScissors also shows robustness against AMC. PrivaScissors achieves attack accuracy of the rate of random guess by sacrificing less than 1\% and 0.5\% model accuracy on CIFAR10 and CIFAR100, respectively. The other baselines achieve the same defense performance by sacrificing more than 5\% and 4\% model accuracy, respectively.

\subsection{Integration of Data and Prediction protection}

\begin{figure}[h]
\centering
     \includegraphics[scale=0.35]{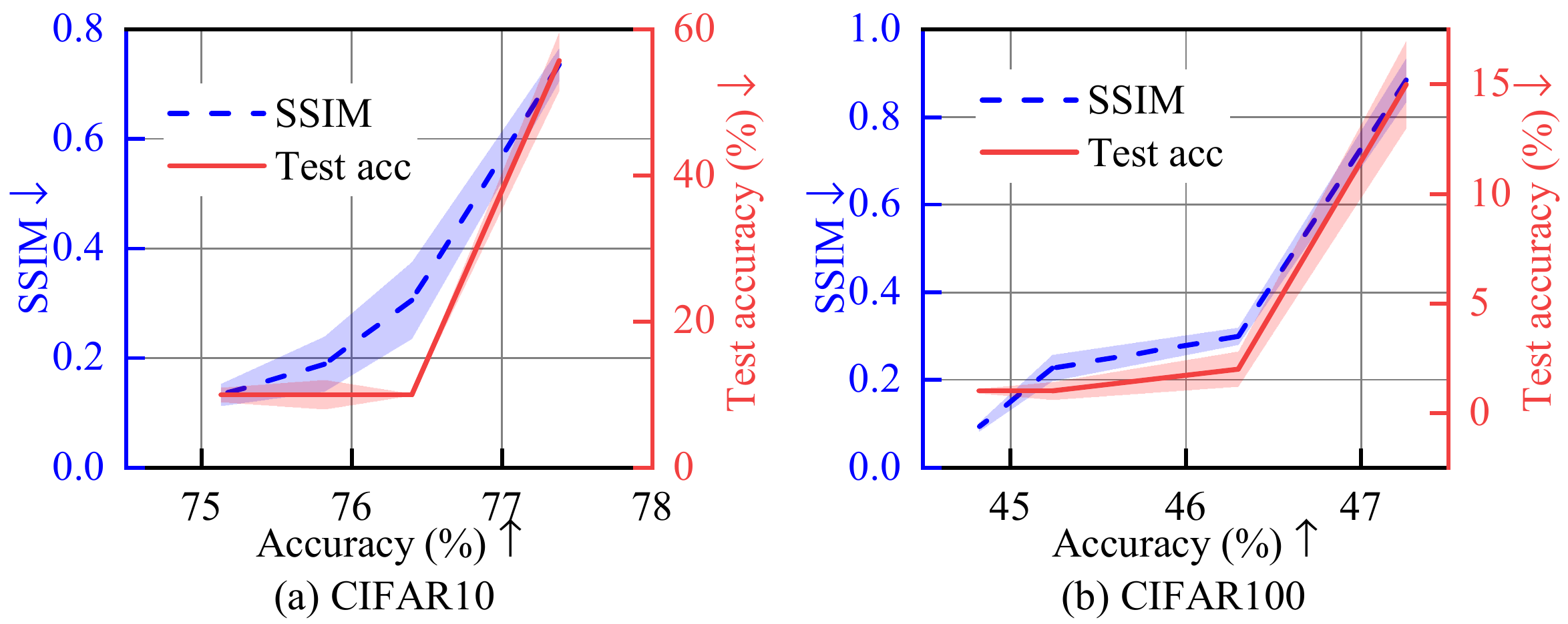}
\caption{PrivaScissors against KA and PMC on CIFAR10 and CIFAR100.}
\label{fig:integrate}
\end{figure}

We have shown the compared results of data and prediction protection between PrivaScissors and the baselines in \cref{sec:results_data} and \cref{sec:results_prediction}. In this section, we evaluate the integration of data and prediction protection of PrivaScissors. We set $\lambda_d$ and $\lambda_l$ between 0.05-0.4 and evaluate the defenses. The results of defense against the KA and PMC attacks on CIFAR10 and CIFAR100 are shown in \cref{fig:integrate}. It is shown that PrivaScissors can effectively protect data and prediction simultaneously with less than a 2\% accuracy drop for both datasets.

\section{Conclusion}

We propose a defense method PrivaScissors, against privacy leakage in collaborative inference by reducing the mutual information between the model's intermediate outcomes and the device's data and predictions. The experimental results show that PrivaScissors can defend against data leakage and prediction leakage effectively. We also provide a theoretically certified robustness guarantee for PrivaScissors. In this paper, we focus on the scenario where there is only one edge device. Our defense can be easily applied to the collaborative inference scenario with multiple edge devices.

\clearpage

\bibliographystyle{ieeetr}
\bibliography{reference}

\clearpage

\appendix
\section{Algorithm}\label{app:algorithm}

\begin{algorithm}[h]  
\footnotesize
\renewcommand{\algorithmicrequire}{\textbf{Input:}}
\renewcommand{\algorithmicensure}{\textbf{Output:}}
    \caption{\textbf{Training algorithm of PrivaScissors.} $\textcolor{green}{\leftarrow}$ means information is sent to the server; $\textcolor{blue}{\leftarrow}$ means information is sent to the device; \textcolor{red}{red steps} are conducted on the cloud server.}
    \label{alg:Framwork}  
    \begin{algorithmic}[1] 
        \Require Dataset $\left\{\left(x_i, y_i\right)\right\}_{i=1}^N$; Learning rate $\eta$. 
        \Ensure $\theta^h; \theta^e; \theta^c; \psi; \phi$.
        \State Initialize $\theta^h, \theta^e, \theta^c, \psi, \phi$;
        \For {a batch of data $\left\{\left(x_i, y_i\right)\right\}_{i\in \sB}$}
            \State $\{r_i\}_{i\in\sB}\textcolor{green}{\leftarrow}\{f^h_{\theta^h}\left(x_i\right)\}_{i\in\sB}$;
            \State $\mathcal{L}_{d\_a}\leftarrow\frac{1}{|\sB|} \sum\limits_{i\in\sB} \log g_{\psi}\left(x_i|r_i\right)$;
            \State $\psi \leftarrow \psi + \eta\nabla_{\psi}\mathcal{L}_{d\_a}$;
            
            \State $\{z_i\}_{i\in\sB}\textcolor{blue}{\leftarrow}\textcolor{red}{\{f^e_{\theta^e}\left(r_i\right)\}_{i\in\sB}}$;
            
            \State $\mathcal{L}_c\leftarrow\frac{1}{|\sB|} \sum\limits_{i\in\sB}\mathcal{L}\left(f^c_{\theta^c}\left(z_i\right), y_i\right)$;
            \State $\mathcal{L}_{l\_a}\leftarrow\frac{1}{|\sB|} \sum\limits_{i\in\sB}\log h_{\phi}\left(y_i|z_i\right)$;
            \State $\theta^c \leftarrow \theta^c - \eta\nabla_{\theta^c}\mathcal{L}_c$;
            \State $\phi \leftarrow \phi + \eta\nabla_{\phi}\mathcal{L}_{l\_a}$;
            
            \State $\{y_{n_i'}\}_{i\in\sB}\leftarrow$ randomly sample $\{y_{n_i'}\}_{i\in\sB}$ from $\{y_i\}_{i\in[N]}$;
            \State $\{x_{k_i'}\}_{i\in\sB}\leftarrow$ randomly sample $\{x_{k_i'}\}_{i\in\sB}$ from $\{x_i\}_{i\in[N]}$;
            \State $\mathcal{L}_{d\_r}\leftarrow \frac{1}{|\sB|} \sum\limits_{i\in\sB} -\log g_{\psi}\left(x_{k_i'}|r_i^2\right)$;
            \State $\mathcal{L}_{l\_r}\leftarrow \frac{1}{|\sB|} \sum\limits_{i\in\sB} -\log h_{\phi}\left(y_{n_i'}|z_i^2\right)$;
            
            \State $\{\nabla_{z_i}\mathcal{L}\}_{i\in\sB} \textcolor{green}{\leftarrow} \{\nabla_{z_i}\left[(1-\lambda_d-\lambda_l)\mathcal{L}_c+\lambda_l\mathcal{L}_{l\_a}+\lambda_l\mathcal{L}_{l\_r}+\lambda_d\mathcal{L}_{d\_a}+\lambda_d\mathcal{L}_{d\_r}\right]\}_{i\in\sB}$;
            
            \State $\textcolor{red}{{\nabla_{\theta^e}\mathcal{L} \leftarrow \frac{1}{|\sB|}\sum\limits_{i\in\sB}}\nabla_{z_i}\mathcal{L}\nabla_{\theta^e}{z_i}}$
            \State $\textcolor{red}{\theta^e \leftarrow \theta^e - \eta\nabla_{\theta^e}\mathcal{L}}$;
            \State $\{\nabla_{r_i}\mathcal{L}\}_{i\in\sB} \textcolor{blue}{\leftarrow} \textcolor{red}{\{\nabla_{z_i}\mathcal{L}\nabla_{r_i}{z_i}\}_{i\in\sB}}$;
            \State ${\nabla_{\theta^h}\mathcal{L} \leftarrow \frac{1}{|\sB|}\sum\limits_{i\in\sB}}\nabla_{r_i}\mathcal{L}\nabla_{\theta^h}{r_i}$;
            \State $\theta^h \leftarrow \theta^h - \eta\nabla_{\theta^h}\mathcal{L}$;
        \EndFor
        
    \end{algorithmic} \label{alg:defense}
\end{algorithm}
\section{Proofs of theorems}\label{app:proof}

\begin{proof}
    According to Corollary 3.3 in \cite{cheng2020club}, we have:
    \begin{equation}
        \text{I}(\rz;\ry) < \text{I}_{\text{vCLUB}}(\rz;\ry) + \text{KL}(p(y|z)||h_{\phi}(y|z)).
    \end{equation}
    Then we have
    \begin{equation}
        \text{I}(\rz;\ry) = \mathbb{E}_{p\left(\rz, \ry\right)}\log p\left(\ry|\rz\right) - \mathbb{E}_{p\left(\ry\right)}\log p\left(\ry\right) < \epsilon,
    \end{equation}
    where $\epsilon=\text{I}_{\text{vCLUB}}(\rz;\ry) + \text{KL}(p(y|z)||h_{\phi}(y|z))$. With the samples $\{x_i,y_i\}$, $\text{I}(\rz;\ry)$ has an unbiased estimation as:
    \begin{equation}
        \frac{1}{N}\sum\limits_{i=1}^N \log p(y_i|z_i) - \frac{1}{N}\sum\limits_{i=1}^N \log p(y_i) < \epsilon.
        \label{eq:proof_prediction}
    \end{equation}
    Suppose the adversary has an optimal model $h^m$ to estimate $p(y_i|z_i)$ such that $h^m(y_i|z_i)=p(y_i|z_i)$ for any $i$, then
    \begin{equation}
        \frac{1}{N}\sum\limits_{i=1}^N \log h^m(y_i|z_i) - \frac{1}{N}\sum\limits_{i=1}^N \log p(y_i) < \epsilon.
    \end{equation}
    For \textbf{classification tasks}, we have
    \begin{equation}
        \frac{1}{N}\sum\limits_{i=1}^N\text{CE}\left[h^m(z_i), y_i\right] > \text{CE}_{random} - \epsilon.
    \end{equation}
    
\end{proof}

\begin{proof}
    Similar with \cref{eq:proof_prediction}, we derive the following for data protection:
    \begin{equation}
        \frac{1}{N}\sum\limits_{i=1}^N \log p(x_i|r_i) - \frac{1}{N}\sum\limits_{i=1}^N \log p(x_i) < \epsilon,
    \end{equation}
    where $\epsilon=\text{I}_{\text{vCLUB}}(\rr;\rx) + \text{KL}(p(x|r)||g_{\psi}(x|r))$. Following the assumption that $p(x|r)$ follows a Gaussian distribution of variance 1, suppose the adversary obtains an optimal estimator $g_m$ of the mean of $p(x|r)$ such that $g^m(x_i|r_i)=p(x_i|r_i)$ for any $i$. Then we have
    \begin{equation}
        \begin{aligned}
        \frac{1}{N}\sum\limits_{i=1}^N \log g^m(x_i|r_i) &< \frac{1}{N}\sum\limits_{i=1}^N \log p(x_i) + \epsilon\\
        \frac{1}{N}\sum\limits_{i=1}^N \log \frac{1}{\sqrt{2\pi}}e^{-\frac{1}{2}\left[x_i-g^m(r_i)\right]^T\left[x_i-g^m(r_i)\right]} &< \frac{1}{N}\sum\limits_{i=1}^N \log p(x_i) + \epsilon\\
        -\frac{1}{N}\sum\limits_{i=1}^N \log \sqrt{2\pi} -\frac{1}{2N}\sum\limits_{i=1}^N \left[x_i-g^m(r_i)\right]^T\left[x_i-g^m(r_i)\right] &< \frac{1}{N}\sum\limits_{i=1}^N \log p(x_i) + \epsilon\\
        \frac{1}{2N}\sum\limits_{i=1}^N \left[x_i-g^m(r_i)\right]^T\left[x_i-g^m(r_i)\right] &> \frac{1}{N}\sum\limits_{i=1}^N \log \frac{\sqrt{2\pi}}{p(x_i)} - \epsilon.\\
        \end{aligned}
    \end{equation}
    We denote the dimension of $\rx$ as Q and $\frac{1}{N}\sum\limits_{i=1}^N \log \frac{\sqrt{2\pi}}{p(x_i)}$ as $\kappa$. Then we have

    \begin{equation}
        \begin{aligned}
            \frac{1}{N}\sum\limits_{i=1}^N \text{MSE}\left[g^m(r_i), x_i\right] > \frac{2(\kappa-\epsilon)}{Q}.
        \end{aligned}
    \end{equation}
    
\end{proof}

\end{document}